\documentstyle[11pt,aaspp4,nick]{article}
\begin{document}


\title{Generation of the Primordial Magnetic Fields during Cosmological
Reionization}
\author{Nickolay Y.\ Gnedin\altaffilmark{1}, Andrea Ferrara\altaffilmark{2}, 
and Ellen G.\ Zweibel\altaffilmark{3}}
\altaffiltext{1}{Center for Astrophysics and Space Astronomy, 
University of Colorado, Boulder, CO 80309;
e-mail: \sl gnedin@casa.colorado.edu}
\altaffiltext{2}{Osservatorio Astrofisico di Arcetri, I-50125 Firenze, Italy;
e-mail: \sl ferrara@arcetri.astro.it}
\altaffiltext{3}{JILA,
University of Colorado, Boulder, CO 80309;
e-mail: \sl zweibel@solarz.colorado.edu}


\load{\scriptsize}{\sc}

\def\A{{\cal A}}
\def\B{{\cal B}}
\def\ion#1#2{\rm #1\,\sc #2}
\def\HI{{\ion{H}{i}}}
\def\HII{{\ion{H}{ii}}}
\def\GI{{\ion{He}{i}}}
\def\GII{{\ion{He}{ii}}}
\def\GIII{{\ion{He}{iii}}}
\def\MH{{{\rm H}_2}}
\def\Hp{{{\rm H}_2^+}}
\def\Hm{{{\rm H}^-}}
\def\etc{{\frenchspacing\it etc. }}
\def\etal{{\it et al.~\/}}
\def\cf{{\it cf.\/}}
\def\ie{{\it i.e.~\/}}
\def\eg{{\it e.g.\/}}
\def\ltsima{$\; \buildrel < \over \sim \;$}
\def\simlt{\lower.5ex\hbox{\ltsima}}
\def\gtsima{$\; \buildrel > \over \sim \;$}
\def\simgt{\lower.5ex\hbox{\gtsima}}

\def\dim#1{\mbox{\,#1}}

\def\figdir{.}
\def\placefig#1{#1}

\begin{abstract}
We investigate the generation of magnetic field by the Biermann battery
 in cosmological
ionization fronts, using new simulations of the reionization of the universe
by stars in protogalaxies. Two mechanisms 
are primarily responsible for magnetogenesis: i) the breakout of I-fronts
from protogalaxies, and ii) the propagation of I-fronts through the high 
density
neutral filaments which are part of the cosmic web. The first mechanism is
dominant prior to overlapping of ionized regions ($z\approx 7$), whereas the 
second 
continues to operate even after that epoch. However, after overlap
the field strength increase is largely due to the gas compression 
occurring as cosmic 
structures form. As a consequence, the magnetic field at $z\approx 5$ closely 
traces 
the gas density, and it is highly ordered on megaparsec scales. The 
mean mass-weighted field
strength is $B_0\approx 10^{-19}\dim{G}$ in the simulation box. There is a
relatively well-defined, nearly linear correlation between $B_0$ and the
baryonic mass of virialized objects, with $B_0\approx 10^{-18}\dim{G}$ in the
most massive objects ($M\approx 10^9 M_\odot$) in our simulations. This is a
lower limit, as lack of numerical resolution prevents us from following small
scale dynamical processes which could amplify the field in protogalaxies. 
Although the field 
strengths we compute are probably adequate as seed fields for
a galactic dynamo, the field is too small to have had significant effects 
on galaxy
formation, on thermal conduction, or on cosmic ray transport in the 
intergalactic medium. It could, however, be observed in the intergalactic
medium through innovative methods based on analysis of $\gamma$-ray burst
photon arrival times.
\end{abstract}

\keywords{cosmology: theory - cosmology: large-scale structure of universe -
galaxies: formation - galaxies: intergalactic medium}

\section{Introduction}

The origin of magnetic fields in the universe is one of the most important 
unsolved problems in cosmology, and has possible ramifications for theories
of galaxy formation and evolution (Wasserman 1978, Kim \etal 1996, Kulsrud
\etal 1997) and for theories of the formation of the first stars (Yamada \&
Nishi 1998, Omukai \& Nishi 1998, Abel \etal 1998, Nakamura \& Umemura 1999,
Bromm, Coppi, \& Larson 1999); see also Rees (1987).

It is widely accepted that
astrophysical magnetic fields reached their present state in a two stage
process; first, the generation of a seed field, and then hydromagnetic
processes which amplified the field or smoothed it.
What are the possible sources for the origin of such
seed field? The question has been posed repeatedly since the original work 
by Biermann (1950). Generally speaking, at least three different possible scenarios have
been put forward. The first one assumes that a magnetic field is present {\it ab
initio}: this does not violate homogeneity but makes the cosmological expansion
anisotropic, and affects primordial nucleosynthesis via an increase of electron
density. The field could have been generated either during QCD/electroweak 
first order cosmic phase
transitions (Quashnock, Loeb \& Spergel 1989, Vachaspati 1991) or during
 inflation 
(Turner \& Widrow 1988, Ratra 1992). 
The upper limits on the present field strength obtained from CMB anisotropy 
measurements (Barrow, Ferreira \& Silk 1997 give 
$B_0 \le 7\times 10^{-9}(\Omega
h^2)^{1/2}\dim{G}$) appear to be more stringent than those derived from the Big
Bang nucleosynthesis. The latter 
have been recently revised by Grasso \& Rubinstein (1995), who found $B_0\le 3\times
10^{-7}\dim{G}$. The upper limit from CMB measurements 
comes very close to the minimum value ($\approx 10^{-9}\dim{G}$)
required for magnetic stresses to 
have been dynamically important (for example in the
creation of voids, de Araujo \& Opher 1997) in the post-recombination 
epoch. Compression alone, without a dynamo,
would have brought such fields to the $\mu$G strengths characteristic of
magnetic fields in galaxies. It is not clear that magnetic fields
generated in the early universe possessed the degree of spatial coherence seen
in galactic magnetic fields (Zweibel \& Heiles 1997) but they could have
been substantially reconfigured by turbulence and differential rotation in
galaxies.

The second type of field generation mechanism  would have operated prior to the
recombination epoch, and was first proposed by Harrison (1970). This process 
invokes
a Biermann battery (originally proposed for stars), 
\ie nonparallel pressure and density gradients, powered by 
a non-zero vorticity in the primordial fluctuation field. 
Electrons and ions would have tended to spin at different rates
due to the drag on electrons by the CMB. An
ambipolar electric field would have arisen to couple the electrons
and ions, and a magnetic field would thus have been created by induction.
Zeldovich, Ruzmaikin, \& Sokoloff (1983) later discussed a variant of this
mechanism. Even if the vorticity is equivalent to galactic rotation by
$z = 10$, the magnitude of the resulting magnetic field is only of order $10^{-21}\dim{G}$. However,
vorticity modes are now thought to decay rapidly as irrotational 
modes grow and structure forms;
thus this process now appears very unlikely.

In the third class of schemes, magnetic fields can arise during the epoch
of protogalaxy formation, through any mechanism which produces a non-potential 
electric field. It is expected that fields produced in this manner 
must be small, 
and only act as a seed for a hydromagnetic dynamo. Two 
mechanisms for producing seed fields have been considered. In the first,
magnetogenesis proceeds primarily in shocks.
Davies \& Widrow (1999) showed with simple analytic models that magnetic 
fields as high as $10^{-17}\dim{G}$ can be generated during the collapse of 
 protogalaxies. 
Kulsrud \etal (1997) computed
the effect of the Biermann battery in a simulation of large scale structure.
They found that magnetic fields are generated primarily in shocks 
associated with
the gravitational collapse of the structure, with a 
a characteristic field strength of $10^{-21}\dim{G}$.

In the second mechanism, magnetogenesis takes place in ionization fronts.
Subramanian, Narasimha, \& Chitre (1994, SNC) pointed out that the 
Biermann battery
would operate in cosmological ionization fronts propagating through density
irregularities. They made simple estimates 
based on the properties of protogalactic fluctuations, and suggested that
fields of a few times $10^{-20}\dim{G}$, with a coherence length of 
several kpc, could be generated by this mechanism. Along similar lines, but
exploiting SN explosions rather than ionization fronts, Miranda, Opher \& Opher
(1998) concluded that field seeds as high as $B\approx 4.5\times 10^{-10}\dim{G}$ can be  
created with coherence scales of order of $100$~kpc. However, their
mechanism requires efficient SN shell fragmentation, 
which is not expected to occur in such low metallicity environments 
(Ferrara 1998). Furthermore,
most of their magnetic flux is produced by objects with mass $\approx 10^6
M_\odot$ forming at redshift $z\approx 300$, which is clearly at odds with all 
currently viable cosmological models.

We should also mention that battery mechanisms could have operated on much 
smaller scales, such as in accretion disks around massive compact objects, and
seeded young galaxies and the IGM through outflows and jets (Daly \& Loeb 
1990).

Observations provide little guidance in choosing among these theoretical
models.
Attempts to determine intergalactic 
magnetic fields have yielded uncertain results, or at best upper limits. 
A very clean determination of primordial $B$-fields could be
accomplished through measurement of 
Faraday rotation in the polarization of the CMB (Kosowsky \& Loeb 1996). 
This will
 become possible with forthcoming experiments.
             
An experimental
upper limit $B_0 \simlt 10^{-9}(\Omega_{IGM}/0.01)\dim{G}$ (see review by
Vall\'ee 1997 and references therein) is found on 
the intergalactic 
$B$-field averaged over a length scale comparable to the horizon. 
This upper bound
would increase by a factor of 3 if the field were 
coherent on $10\dim{Mpc}$ scales (Kronberg 1994).

Finally, there is some evidence for $\mu$G magnetic fields in young galaxies, 
as traced by high redshift (up to $z\approx
3$) quasars (Kronberg, Perry \& Zukowsky 1990) and
damped Ly$\alpha$ systems (Wolfe, Lanzetta \& Oren 1992). 
This value might reflect the saturated
level dictated by equipartition with turbulent energy (Field 
1995), but it requires a rather high initial seed field value in
order for the dynamo to amplify the field up to that strength.    
At that epoch, a dynamo working in a galaxy like the Milky Way
should be provided a seed $B \simgt 10^{-18}-10^{-17}\dim{G}$
(Beck \etal 1996).

The purpose of this paper is to explore magnetogenesis in cosmological
ionization fronts using new simulations of the reionization of the universe
by stars in protogalaxies. Our results include, in a self consistent manner,
the wealth of density structures present at the epoch of galaxy formation
(at least down to a scale imposed by numerical resolution). We corroborate
the results of SNC to order of magnitude, 
place their suggestion on a firmer ground, isolate the most
important mechanisms, and calculate the structure of the magnetic field in 
greater detail.

In \S 2 we describe the simulations, \S 3 gives the results, and \S 4 is a
summary of the calculations and their limitations, as well as a discussion of
their astrophysical relevance.

\section{Method}

\subsection{Simulations}

We use cosmological hydrodynamic simulations performed with the 
``Softened Lagrangian Hydrodynamics'' (SLH) code and reported
in Gnedin (2000). These simulations include 3D radiative
transfer (in an approximate implementation) and are therefore
suitable for following magnetic
field generation. In addition, the simulations include a
phenomenological description
of the star formation process, based on the Schmidt law and on a
full time-dependent 
treatment of atomic and molecular physics in a plasma with primeval
composition.

\def\tableone{
\begin{deluxetable}{ccccc}
\tablecaption{Numerical Parameters\label{tabone}}
\tablehead{
\colhead{Run} & 
\colhead{$N$} & 
\colhead{Box size} & 
\colhead{Baryonic mass res.} & 
\colhead{Spatial res.} }
\startdata
A & $128^3$ & $4h^{-1}{\rm\,Mpc}$ & $10^{5.7}\dim{M}_{\sun}$ & 
$1.0h^{-1}{\rm\,kpc}$ \\
B & $64^3$ & $2h^{-1}{\rm\,Mpc}$ & $10^{5.7}\dim{M}_{\sun}$ & 
$1.5h^{-1}{\rm\,kpc}$ \\
C & $128^3$ & $2h^{-1}{\rm\,Mpc}$ & $10^{4.8}\dim{M}_{\sun}$ & 
$0.5h^{-1}{\rm\,kpc}$ \\
\enddata
\end{deluxetable}
}
\placefig{\tableone}
We adopt a currently fashionable variant of the
CDM+$\Lambda$ cosmological model with $
\Omega_0 = 0.3$, $\Omega_\Lambda = 0.7$, $h = 0.7$, and
$\Omega_b = 0.04$. We have run three simulations with varying resolution,
to estimate the level of numerical convergence. The numerical parameters
of all three runs are listed in Table \ref{tabone}.
Runs A and B are stopped at $z=4$
because at lower redshift the simulation box cannot be considered as
a representative region of the universe: in fact, the rms density perturbation
on the scale of the box size exceeds 0.25. Run C was stopped at $z=6.5$
because it is used for estimating numerical convergence only.

We will use run A as our reference run, whereas runs B and C
will serve as a measure of the theoretical uncertainty of
our calculations.

The placement of the simulation box of run A
is designed in such a way as to marginally resolve the characteristic 
filtering scale below
which the baryonic perturbations are smoothed due to the finite
gas pressure. Thus, run A includes 
essentially all (about 93\%) of the small scale power that is present in the 
initial conditions (Gnedin 2000). 
Since the small scale perturbations can contribute
significantly to the total production rate of the magnetic field, we thus
ensure that we do not miss a substantial portion of the total production
in the regions with moderate overdensities. However, inside the virialized
objects small scale power can be generated gravitationally, and since
this power is not resolved, we are almost certainly underestimating the
true value of the magnetic field. We argue below, however, that the magnetic 
field on large scales is generated primarily by structure on large scales.
Thus, our calculation should produce a good approximation to the large scale
fields produced by magnetogenesis in ionization fronts. The next
stage, amplification of
these primordial fields by dynamo action inside virialized objects, is
outside the scope of this study. 

\subsection{Magnetic Field Equation}

The evolution equation for the magnetic field with the Biermann battery
source term can be written as follows:
\begin{equation}
	{\partial \vec{B}\over\partial t} +
	(\vec{v}\cdot \nabla)\vec{B} = -\vec{B} \dim{div}\vec{v} + 
	(\vec{B}\cdot \nabla)
	\vec{v} + {c k_B\over n_e e}\nabla T\times \nabla n_e +
	{{m_ec}\over {e}}\nu_{e\gamma}\nabla\times \vec{v}
	\label{beq}
\end{equation}
where $n_e$ is the electron number density, $T$ is gas temperature,
and 
$$
	\nu_{e\gamma} = {{4\sigma_T c U_{\gamma}}\over {3m_ec^2}},
$$
with $U_{\gamma}$ being the energy density of the CMB, and $\sigma_T$
the Thompson cross section.

The left hand side of equation (\ref{beq}) describes a simple advection
of the magnetic field together with the fluid. The two first terms on the
right hand side of the field (induction) equation can be rewritten in 
index
notation in the following form:
$$
	-B^i v^j_{,j} + B^j v^i_{,j} =
	-{2\over 3}B^i v^j_{,j} + B^j (v^i_{,j}-{1\over3}\delta^i_j v^k_{,k})
	= 
	{2\over 3}B^i {D\rho\over dt} + B^j \sigma^i_j,
$$
where $D\rho/dt$ is the Lagrangian derivative of the gas density, and
$\sigma^i_j$ is the traceless part of the velocity tensor. These two terms
describe two physically different processes of amplification of the
magnetic field : compression and stretching respectively. We discuss
the importance of those two terms below.
The third term in equation (\ref{beq}) is the Biermann battery, and the
last term is due to Compton drag of CMB photons on the free electrons. We ignore the Hall
effect and consider ions and neutrals moving with the same fluid velocity
$\vec{v}$.

From general physical considerations, we expect that 
the battery term always dominates in ionization fronts, and is small
everywhere else. It is useful to estimate the magnetic field produced by the
battery term as an ionization front sweeps through the gas at speed $v_f$. 
Integrating equation (\ref{beq}) in time, keeping only the battery term and
assuming that $\nabla n_e$ is in the direction of propagation of the front
while $\nabla T$ is 
perpendicular to $\nabla n_e$ and is
 produced by some large scale inhomogeneity, we find
$$
B\sim {{ck_BT}\over {ev_f l_g}},
$$
where $l_g$ is gradient length scale for $T$. In order to get some physical
feeling for this field strength, we introduce the ionization time $t_i\equiv
l_g/v_f$, the ion gyrofrequency $\omega_{ci} \equiv eB/m_ic$, and the ion
thermal velocity $v_{th}\equiv (k_BT/m_i)^{1/2}$. In terms of these quantities,
$$
\omega_{ci}t_i\sim{{v_{th}^2}\over {v_f^2}}.
$$
Or, in terms of the thermal ion gyroradius $r_i\equiv v_{th}/\omega_{ci}$,
$$
r_i/l_g\sim{{v_f}\over v_{th}}.
$$
In contrast to the usual astrophysical situation, in which $\omega_{ci}^{-1}$
and $r_i$ are much smaller than any global time scales or length scales, the
battery produces a field so weak that the magnetization scales and global
scales are comparable. But note that small scale structure (small $l_g$)
produces stronger fields. 

The ratio of the photon drag term to the induction term,
assuming that the vorticity is appreciable,
is
\begin{equation}
\sim {{\nu_{e\gamma}}\over {\omega_{ce}}},
\end{equation}
where $\omega_{ce}\sim 10^7 B\dim{s}^{-1}$ is the electron gyrofrequency. 
Since $\nu_{e\gamma}= 5\times 10^{-20}(1+z)^4 \dim{s}^{-1}$,
the induction dominates photon
drag for
\begin{equation}
	B > 5\times 10^{-27} (1+z)^4 \dim{G}.
	\label{cdlim}
\end{equation}

Using the outputs from the simulations, we solved equation (\ref{beq})
numerically using clouds-in-cells (Efstathiou \etal 1985)
to advect the right hand side of
equation (\ref{beq}) over the SLH quasi-Lagrangian mesh, and implementing
the Lax averaging for the stretching term to ensure the stability of the
scheme. We chose the time-step so as to ensure numerical convergence
with respect to time discretization.

We ignore the back reaction of the magnetic field because the magnetic
pressure for the field strength we obtain ($B\sim10^{-19}\dim{G}$, 
$B^2/8\pi\sim 10^{-40} \dim{erg}/\dim{cm}^3$) is far smaller than 
the gas pressure even in the low density IGM ($P\sim 10^{-16}
\dim{erg}/\dim{cm}^3$ at $z\sim7$).

\section{Results}

\subsection{General Outlook}

\def\capEV{
Evolution of the mass ({\it bold lines\/}) and volume ({\it thin 
lines\/}) weighted mean magnetic field strength 
({\it top panel\/}) and the comoving mean free path to ionizing radiation
({\it bottom panel\/})
for
runs A ({\it solid lines\/}), B ({\it dotted lines\/}), and C
({\it dashed lines\/}).
}
\placefig{
\begin{figure}
\insertfigure{\figdir/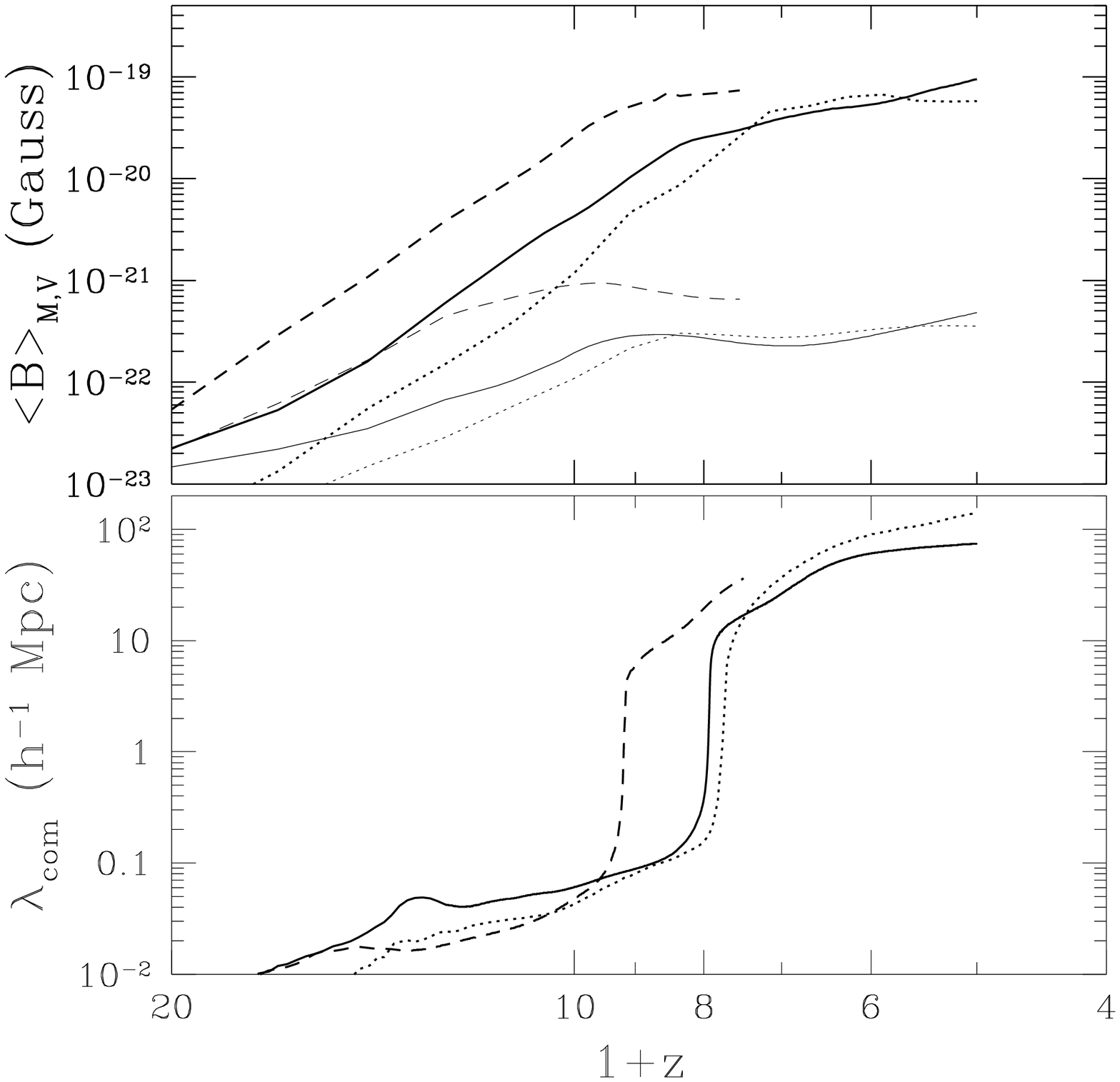}
\caption{\label{figEV}\capEV}
\end{figure}
}
We can now look at general features of the magnetic field generated
during the reionization. Figure \ref{figEV} shows the evolution of the
mass and volume weighted mean magnetic field strength
as well as the mean free path to ionizing radiation  for all three
simulations discussed in this paper.
The difference between the curves illustrates the level of numerical
convergence achieved in the simulations, and can be considered the 
theoretical error of our simulations. In particular, the differences
at high redshift are due to the fact that our reference run A simulation
still misses some 7\% of the small scale power, which translates into
a $\approx 15$\% shift in redshift. However, the saturation levels of all
three runs are about the same.
Since run C has a twice smaller cell size (and,
therefore, approximately four times less numerical diffusion), 
the agreement in the saturated level of the produced magnetic field
demonstrates that numerical diffusion
does not lead to a substantial
underestimate of the magnetic field strength produced in our simulations.
This test is not however completely conclusive because of the
excess small scale power present in run C, and therefore we
cannot exclude the possibility that numerical diffusion affects our
results on a several tens of percent level. However, since we focus in
this paper on the general semi-qualitative description of 
the generation of the primordial magnetic field, we can tolerate a
factor of two uncertainty in our calculations.

We point out here that the
mean field strength generated in our simulations exceeds the value
found by Kulsrud \etal (1997) by about two orders of magnitude. We will
elaborate on this difference in the next subsection.

The evolution of the mean field clearly exhibits two distinct regimes: before
the overlap of the $\HII$ regions at $z=7$ (characterized by a sharp
increase in the mean free path of ionizing photons)
the mean field grows with time, but after the overlap
its growth slows down appreciably. 
\def\capEC{
Evolution of the mass ({\it bold lines\/}) and volume ({\it thin 
lines\/}) weighted mean magnetic field strength 
in the full calculation ({\it solid lines\/}), with the
Compton drag term omitted ({\it long-dashed line\/}), 
with the stretching and the Compton drag
terms omitted ({\it short-dashed line\/}), and with 
the compression and the Compton drag terms omitted
({\it dotted line\/}).
}
\placefig{
\begin{figure}
\epsscale{0.65}
\insertfigure{\figdir/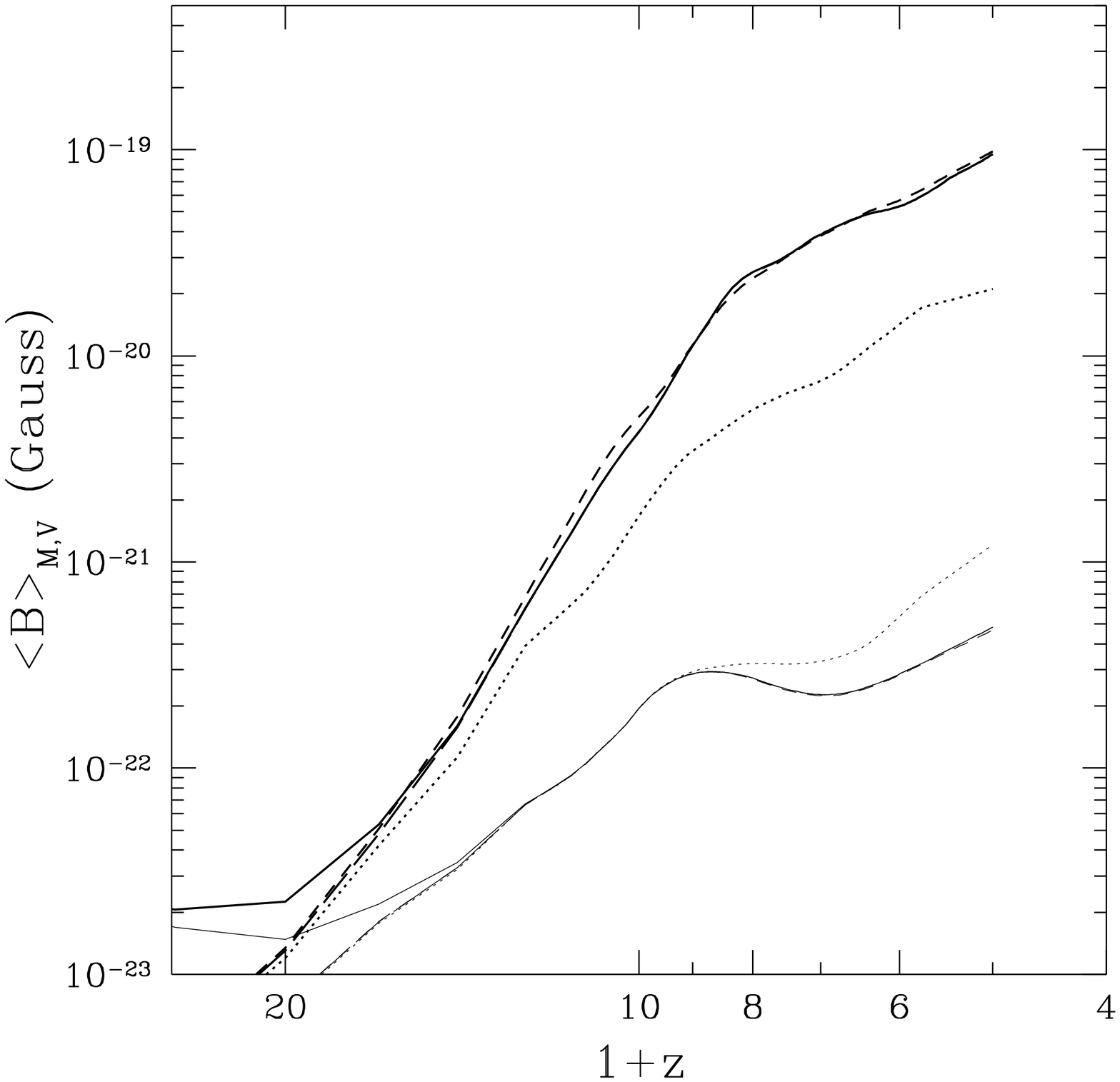}
\caption{\label{figEC}\capEC}
\end{figure}
}
We attempt to understand this behavior by considering which terms in 
equation (\ref{beq}) are dominant at a given time. 
Figure \ref{figEC} shows four solutions to equation (\ref{beq}) for the
large simulation: the solid
lines mark the full solutions (either mass- or volume-weighted), 
the same one as shown in the previous Figure;
the long-dashed line shows a solution with the Compton drag term omitted;
the dashed line shows a solution with both the Compton drag term and
the stretching term omitted; and the
dotted line shows a solution with both the Compton drag term and
the compression term omitted.
We can
immediately conclude that stretching by itself makes an insignificant
contribution, 
whereas compression is
the dominant term after the overlap of
the $\HII$ regions, and the Biermann battery is responsible
for the initial growth of the field during the pre-reionization stage.
The Compton drag term is only important for $z>16$, when
the magnetic field is less than about $5\times10^{-23}\dim{G}$. 
Using the estimate (\ref{cdlim}) we find that at this moment the induction term
is about 10\% of the Compton drag, and the comparison between the dotted and
solid lines in Fig.\ \ref{figEC} shows that the Biermann battery at that time
is on average about 10 times more important than the induction term, in
agreement with our estimate (\ref{cdlim}).

\def\capBD{
The joint mass-weighted 
distribution of the gas density and the magnetic field strength
at $z=4$.
}
\placefig{
\begin{figure}
\epsscale{0.65}
\insertfigure{\figdir/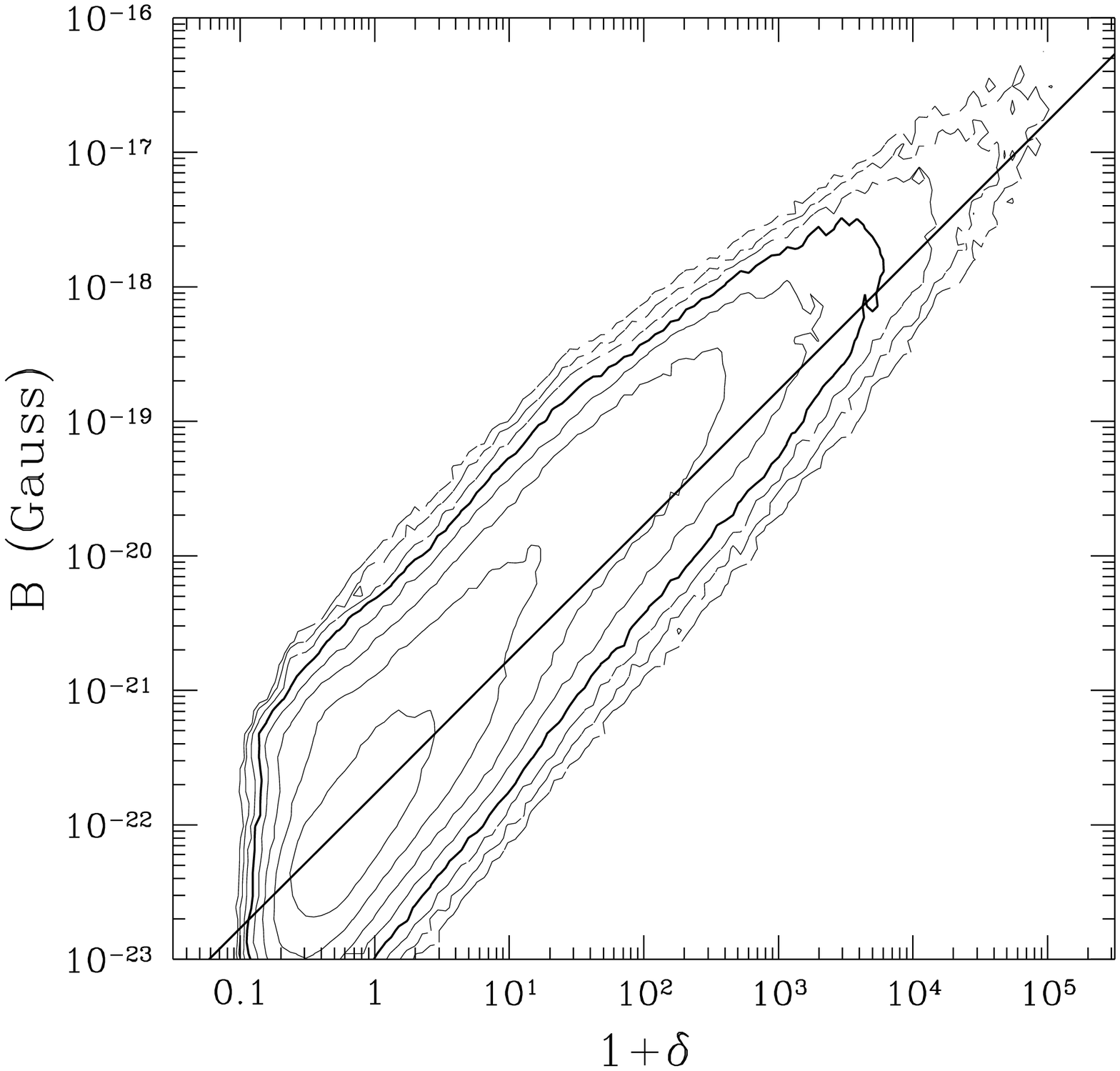}
\caption{\label{figBD}\capBD}
\end{figure}
}
Since compression is the dominant term at lower redshifts, we 
expect that the magnetic field has to be closely related to the
gas density. The battery term has a similar effect, as ionization fronts tend
to move slowly in regions of high gas density.
Figure \ref{figBD} illustrates this point. In this
Figure we show the joint mass-weighted distribution of the gas density
and the magnetic field strength for all fluid elements in the large
simulation. As one can see, there is a strong correlation between the gas 
density and the magnetic field, albeit with a considerable (two orders of 
magnitude in magnetic field strength) scatter. The most probable value for
the magnetic field strength (the peak of the distribution) rises with density
slightly more steeply than the first power. We will give an explanation for this
feature in the next subsection, where we discuss the main mechanism
for generating the primordial magnetic field.

\def\capIM{
A thin slice through the simulation volume at $z=4$, showing in four
panels 
the logarithm of the neutral hydrogen ({\it upper-left\/}) and
the gas density
({\it lower-left\/}) (in units of the average density of the universe),
the gas temperature ({\it lower-right\/}), and
the magnetic field strength ({\it upper-right\/}).
}
\placefig{
\begin{figure}
\insertfigure{\figdir/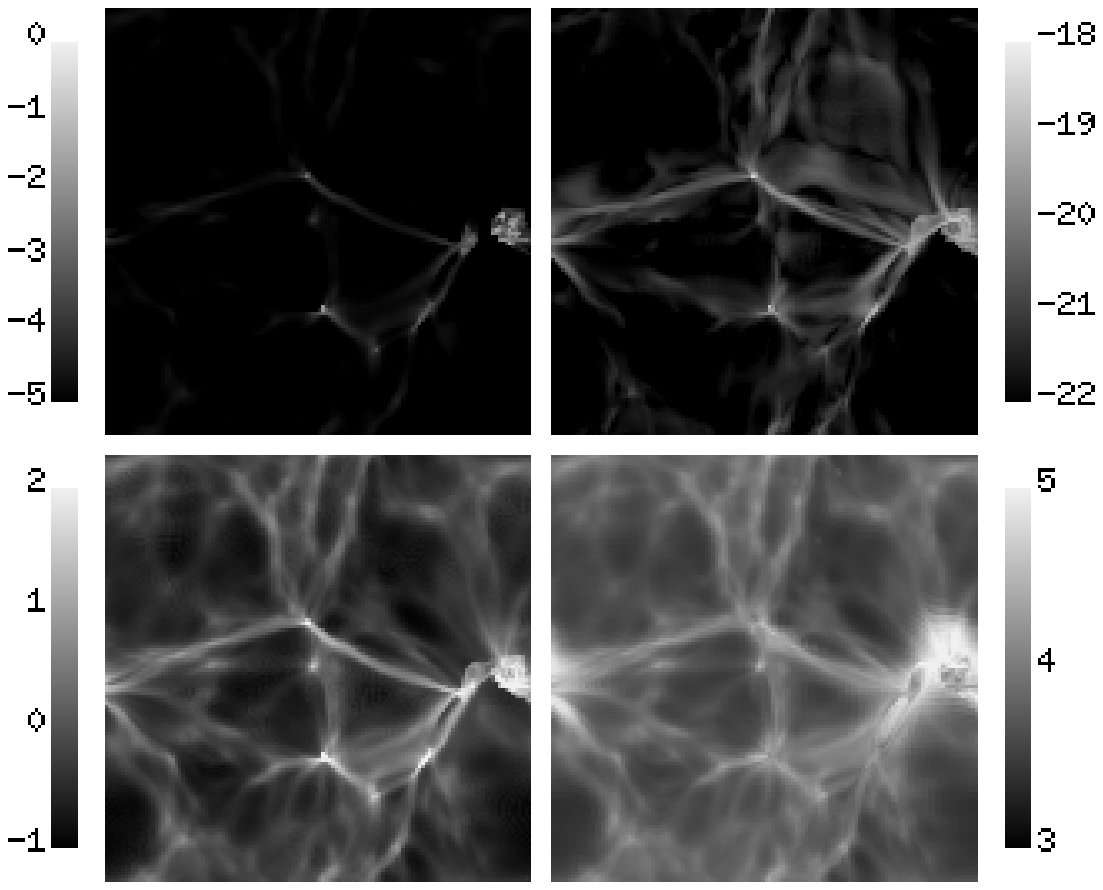}
\caption{\label{figIM}\capIM}
\end{figure}
}
In Figure \ref{figIM} we show a thin slice through the simulation box taken
at a random place. In four panels we show the gas density, temperature,
neutral fraction, and the magnetic field strength (all stretched 
logarithmically).\footnote{Additional visualization of the results of the
simulations including MPEG movies
can be found at the following URL: 
{\tt http://casa.colorado.edu/$\tilde{{\rm ~}}$gnedin/GALLERY/magfi\_p.html}.}
Again, visual inspection
demonstrates that the magnetic field is strongly correlated with the gas
density. It is also correlated with the gas neutral fraction and the
temperature, as these quantities also correlate with the gas density in
the moderate density regime (cosmic overdensity $\delta\la10$).

\def\capIN{
A thin slice through the simulation volume at $z=4$, showing 
the magnetic field strength in the upper-right panel (just as in
Fig.\ \protect{\ref{figIM}}) and three components of the magnetic
field in the other three panel.
}
\placefig{
\begin{figure}
\insertfigure{\figdir/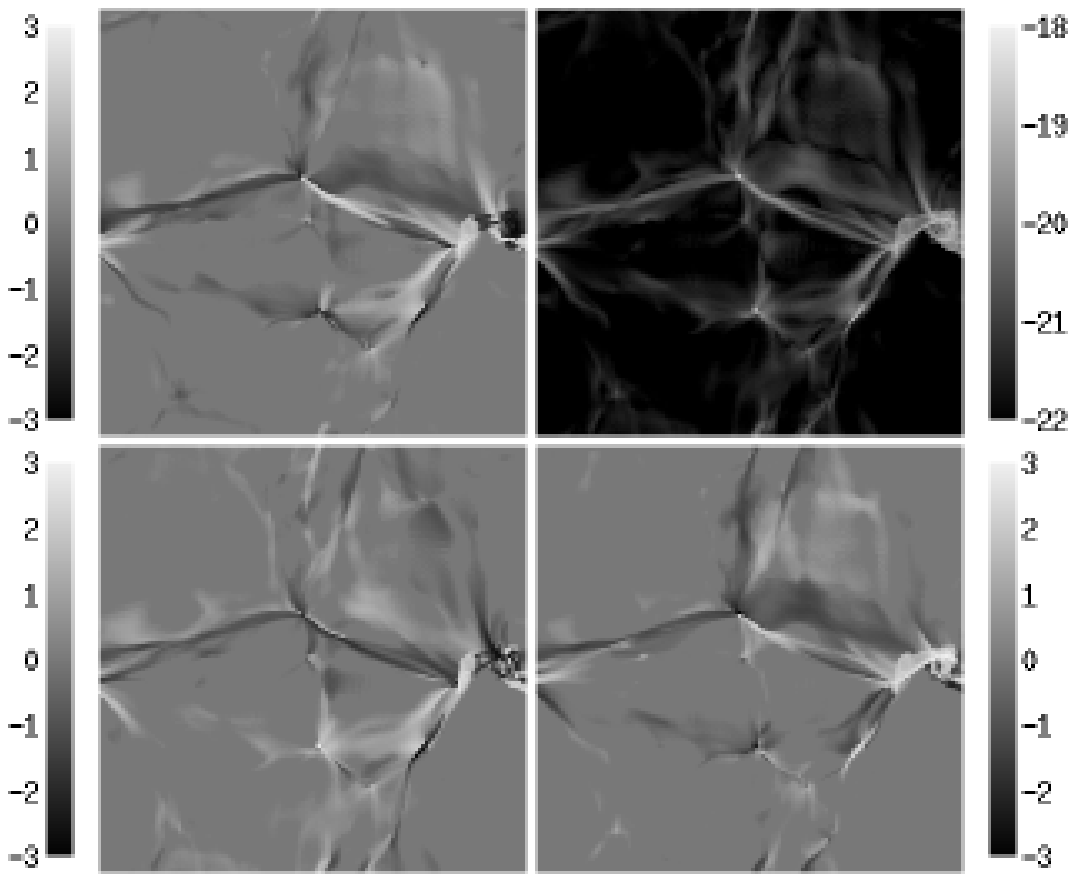}
\caption{\label{figIN}\capIN}
\end{figure}
}
Finally, in Figure \ref{figIN} we show the magnetic field strength
again (just as in Fig.\ \ref{figIM}) together with the three
cartesian components of the magnetic field. Since the components
can be both positive and negative, we use the following stretch to map
the value of the component in the range from $-10^{-19}\dim{G}$ to
$+10^{-19}\dim{G}$ to a quantity in the range from $-3$ to $+3$:
$$
	{\rm stretch}=\dim{sign}(B_i)\dim{log}(1+10^{22}\dim{G}^{-1}|B_i|).
$$
As one can see, the magnetic field is highly ordered on scales of the
order of $1\dim{Mpc}$, and magnetic structure is highly correlated with
density structure. As pointed out above, this is the result of the 
compression occurring as cosmic structures form.
As we have already discussed, this is not an artifact
of  finite numerical resolution, since our simulation resolves all the
relevant scales in the moderate density regime. Only inside the virialized
clumps do we encounter scales which we are unable to resolve, and there
our calculation of the magnetic field strength becomes a lower limit,
on top of which small scale dynamo processes will
generate additional field on galactic and subgalactic scales. Since these
latter processes cannot be simulated reliably with the current resolution
of our simulations, we cannot consider them in this paper. On the other
hand, the extragalactic magnetic fields are calculated reliably in our
simulations for a given cosmological model.

\subsection{Main Mechanisms for Generating the Magnetic Field}

Since the Biermann battery is most efficient when the gradient of the
electron density is perpendicular to the temperature gradient, we can
identify two main mechanisms where the battery is most efficient. But
before we can do so, let us briefly describe the main stages of reionization
(Gnedin 2000). We also point out that the reionization process is affected by
several feedback mechanisms which are discussed in Ciardi \etal (2000).

The reionization
starts with ionization fronts propagating from 
proto-galaxies located in high
density regions into the voids, leaving the high density outskirts of 
the protogalaxies
still neutral, because at high density the recombination time is very 
short, and there are not enough photons to ionize the high density regions.
As an ionization front expands into previously neutral material, it 
leaves behind high density regions whose ionization requires more photons 
than currently available. This stage
of the reionization process can be called ``pre-overlap'', and it extends
over a considerable range of redshifts $\Delta z\sim 5$ around $z\sim 10$.
During this time the high density regions close to the source slowly 
become ionized, whereas more distant high density regions remain neutral.

By $z\approx 7$ (for the cosmological model under consideration)
the $\HII$ regions start to overlap. As a result, an average place in the 
universe can see an increasing number of sources, and the ionizing intensity 
starts to rise rapidly.
The process of reionization enters its second stage, the ``overlap'',
which is quite rapid ($\Delta z\sim 1$).
As the ionizing intensity is rapidly increasing,
the last remains of the neutral low density IGM are quickly eliminated,
the mean free path increases by some two orders of magnitude 
over a Hubble time or so (see Fig. 1), and voids become highly ionized
(neutral fraction of the order of $10^{-5}$). The high density regions at
this moment are still neutral, as the number of ionizing photons available
is not sufficient to ionize them.

After the overlap is complete, the universe is left with both highly ionized 
low 
density regions and some of the high density regions
(which happened to lie close to
the source, where the local value of the ionizing intensity is higher
than the background). High density regions far from any source
remain neutral. This stage can be called ``post-overlap''.
As time goes on, and more and more ionizing photons are emitted, the high 
density regions are gradually eaten away.

\def\capSO{
A cartoon, illustrating the mechanism for generating the
primordial magnetic field during the break through of the I-front
from the proto-galaxy before the overlap of the $\HII$
regions. Enclosed contours show the regions of progressively higher density.
}
\placefig{
\begin{figure}
\epsscale{0.65}
\insertfigure{\figdir/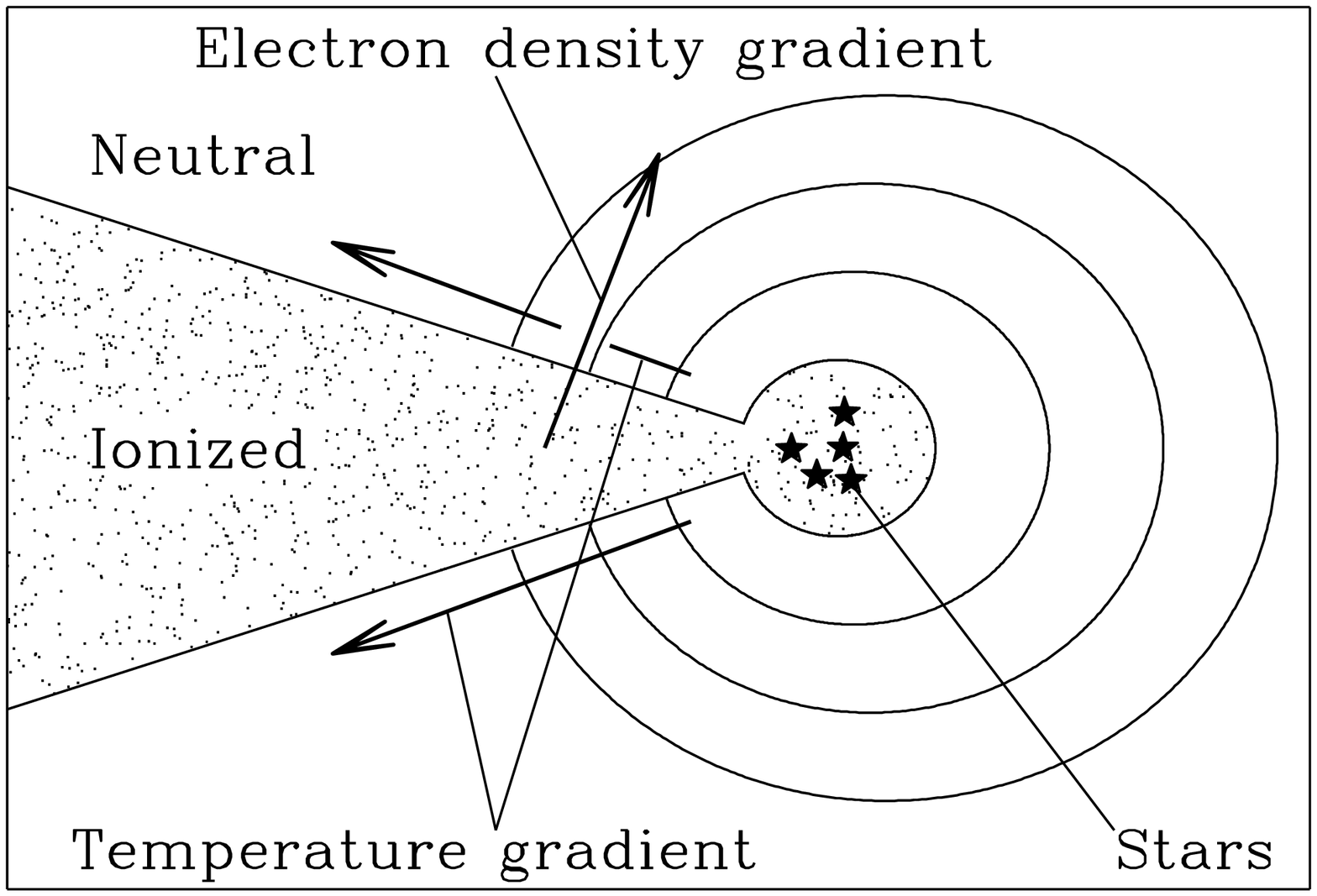}
\caption{\label{figSO}\capSO}
\end{figure}
}
Figure \ref{figSO} shows now a cartoon version of the first out of
two most important 
mechanisms for generating the primordial magnetic field. The 
mechanism takes place when the first I-front breaks through the
protogalaxy. This happens first in a narrow range of angles, where
the gas density happens to be the lowest. As the I-front propagates
into the low density IGM, the gradient of the electron density 
(the surface of the I-front) becomes approximately orthogonal to the
temperature gradient, and the battery becomes maximally efficient.

\def\capIO{
A thin slice through the simulation volume similar to 
Fig.\ \protect{\ref{figIM}} but showing $1/5$th of the box
($0.8h^{-1}\dim{Mpc}$ on a side), aimed at illustrating the generation of
the magnetic field during the break-through of the I-front. Two
redshifts are shown: $z=11.5$ ({\it a\/}) and $z=10.1$ ({\it b\/}).
The region around the I-front is emphasized by the white circle.
}
\placefig{
\begin{figure}
\inserttwofigures{\figdir/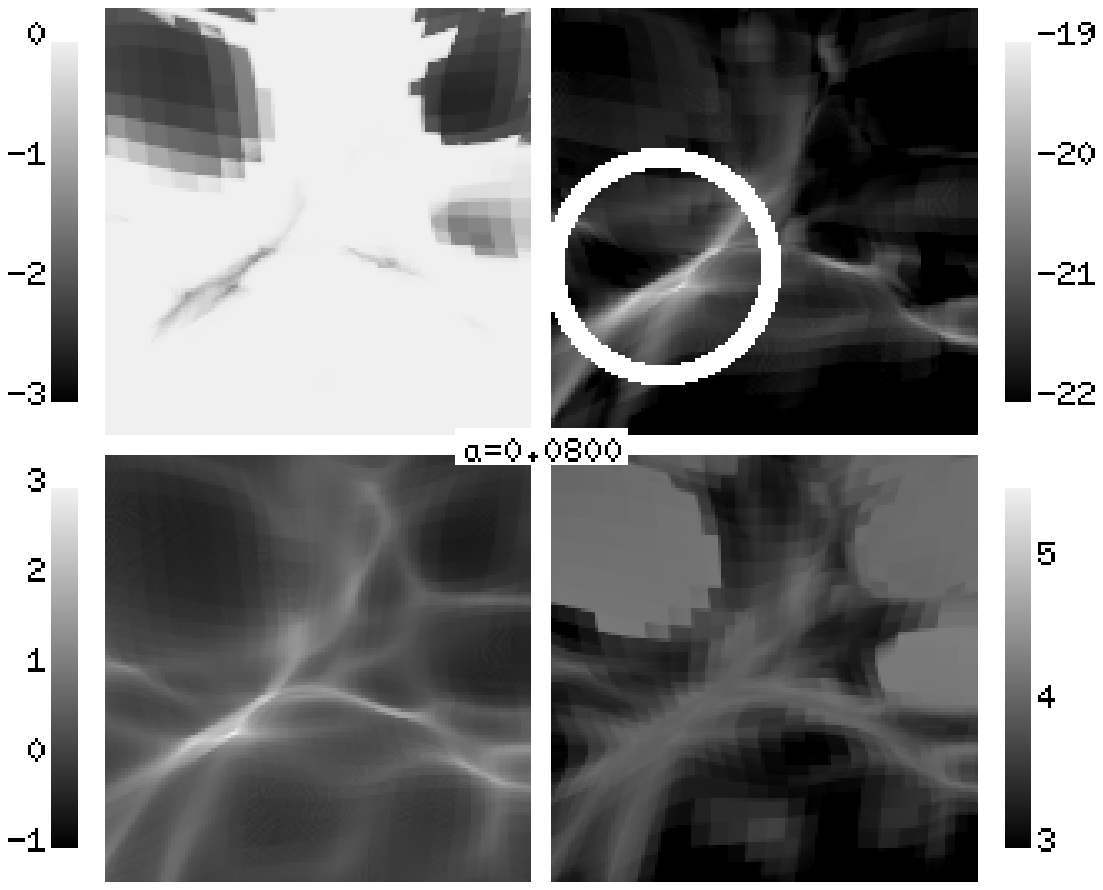}{\figdir/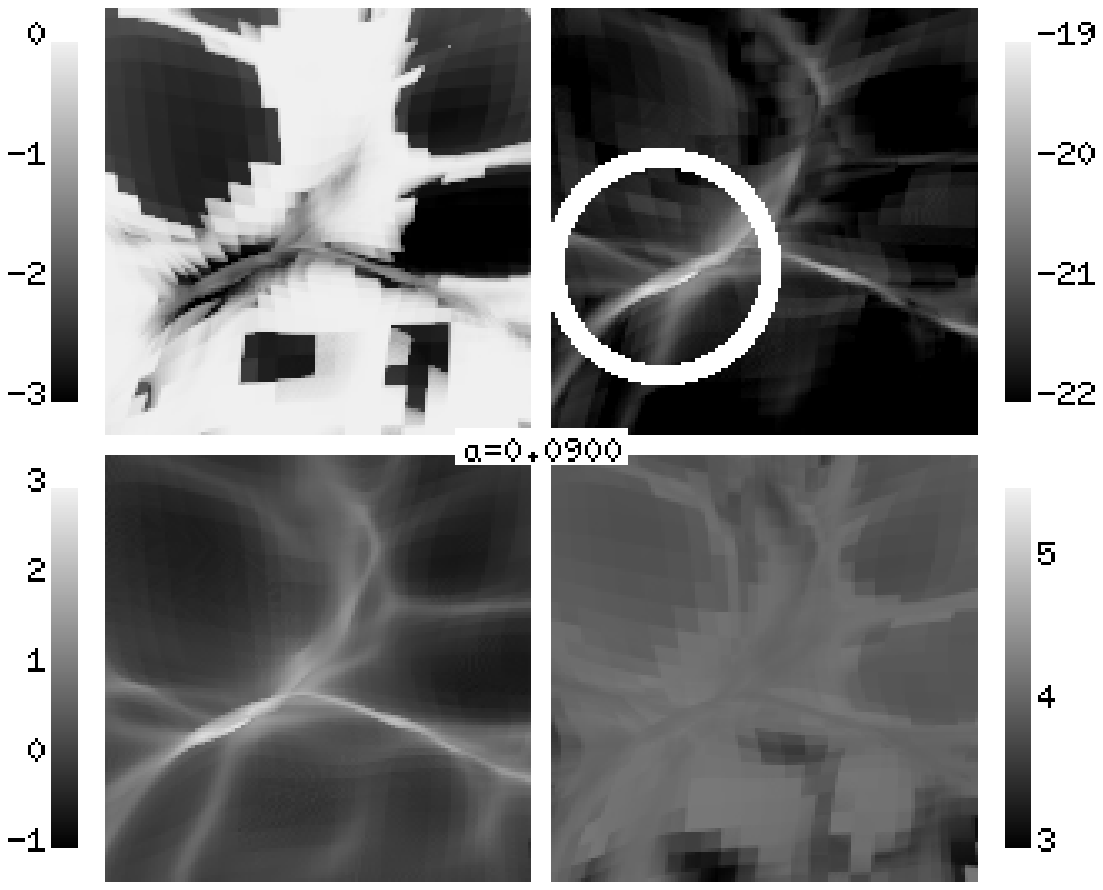}
\caption{\label{figIO}\capIO}
\end{figure}
}
Figure \ref{figIO} illustrates this mechanism with an example from the
simulation. Each of the frames is similar to Fig.\ \ref{figIM},
except that it is only one fifth the box size ($0.8h^{-1}\dim{Mpc}$)
on a side.
Two panels show a protogalaxy just before and after the I-front
broke through the gaseous halo.\footnote{The slice through the simulation box 
is chosen randomly, and is not passing through the center of the object.
We are therefore seeing a cross section of Fig.\ \ref{figSO} taken off
center and at an angle with respect to the axis of the ionized cone.}
The magnetic field has increased by almost
an order of magnitude during the short time interval ($\Delta t= 75
\dim{Myr}$) between the two snapshots.

The magnetic field produced by this mechanism has toroidal
topology, wrapping around the $\HII$ region. This can be seen by inspection
of the battery term: $\nabla n_e$ is directed across the ionization front,
while $\nabla T$ follows the large scale stratification of the protogalaxy
and is directed along the major axis of the cone shaped H II region. The 
magnetic field is oriented along the cross product of these gradients, i.e.
the field encircles the H II region. The reversal of the out-of-plane
component of the field across the H II region
is indeed observed
in the simulation.

\def\capSF{
A cartoon, illustrating the second mechanism for generating the
primordial magnetic field during reionization: the I-front crossing a neutral
high density filament after the epoch of overlap.
Enclosed contours show the regions of progressively higher density.
}
\placefig{
\begin{figure}
\insertfigure{\figdir/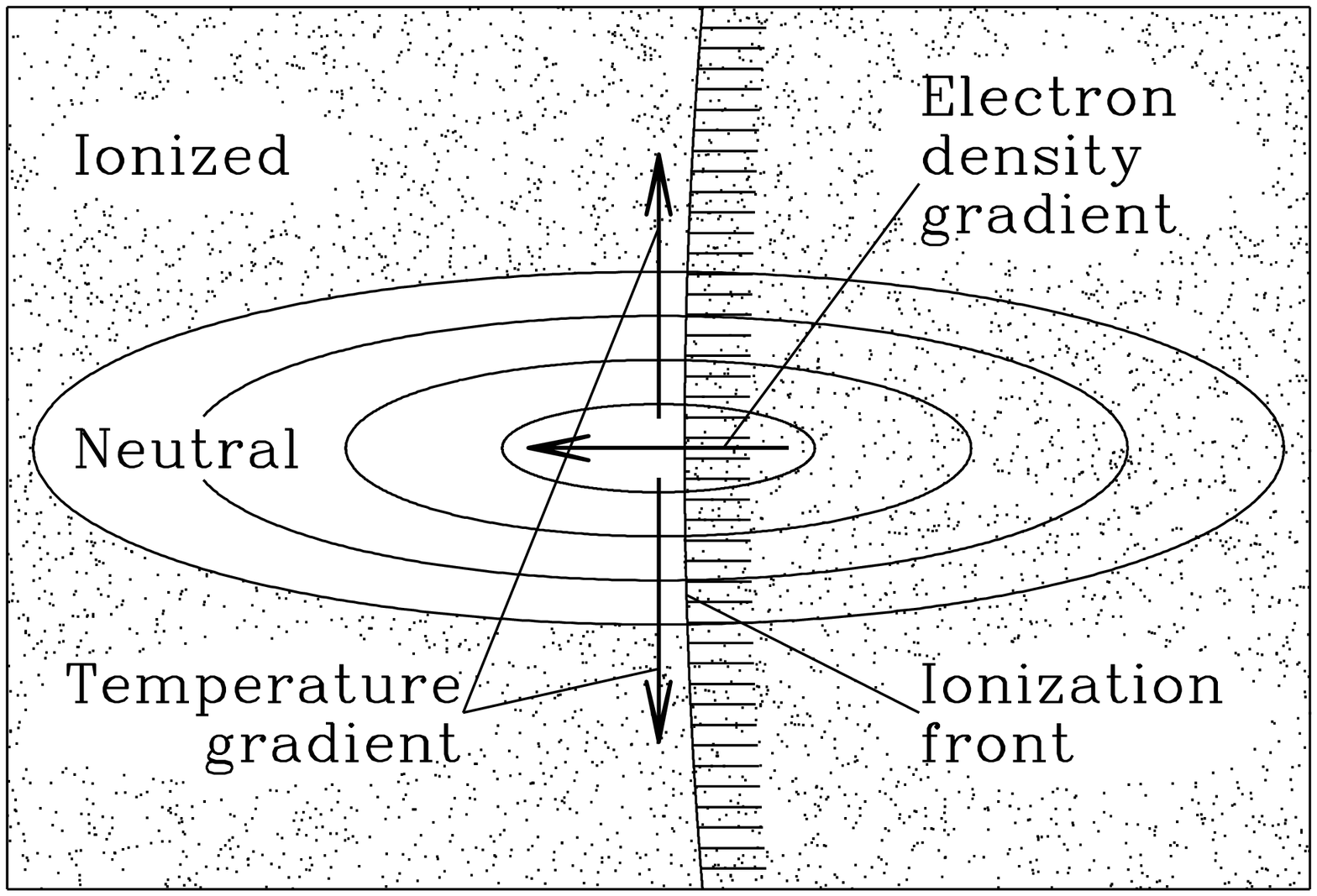}
\caption{\label{figSF}\capSF}
\end{figure}
}
\def\capIF{
A thin slice through the simulation volume similar to 
Fig.\ \protect{\ref{figIM}} but showing $1/25$th of the box
($0.8h^{-1}\dim{Mpc}$ on a side), illustrating the generation of
the magnetic field during the propagation of the I-front across
the high density filament. Tho
redshifts are shown: $z=5.9$ ({\it a\/}) and $z=5.5$ ({\it b\/}).
The region around the I-front is emphasized by the white circle.
}
\placefig{
\begin{figure}
\inserttwofigures{\figdir/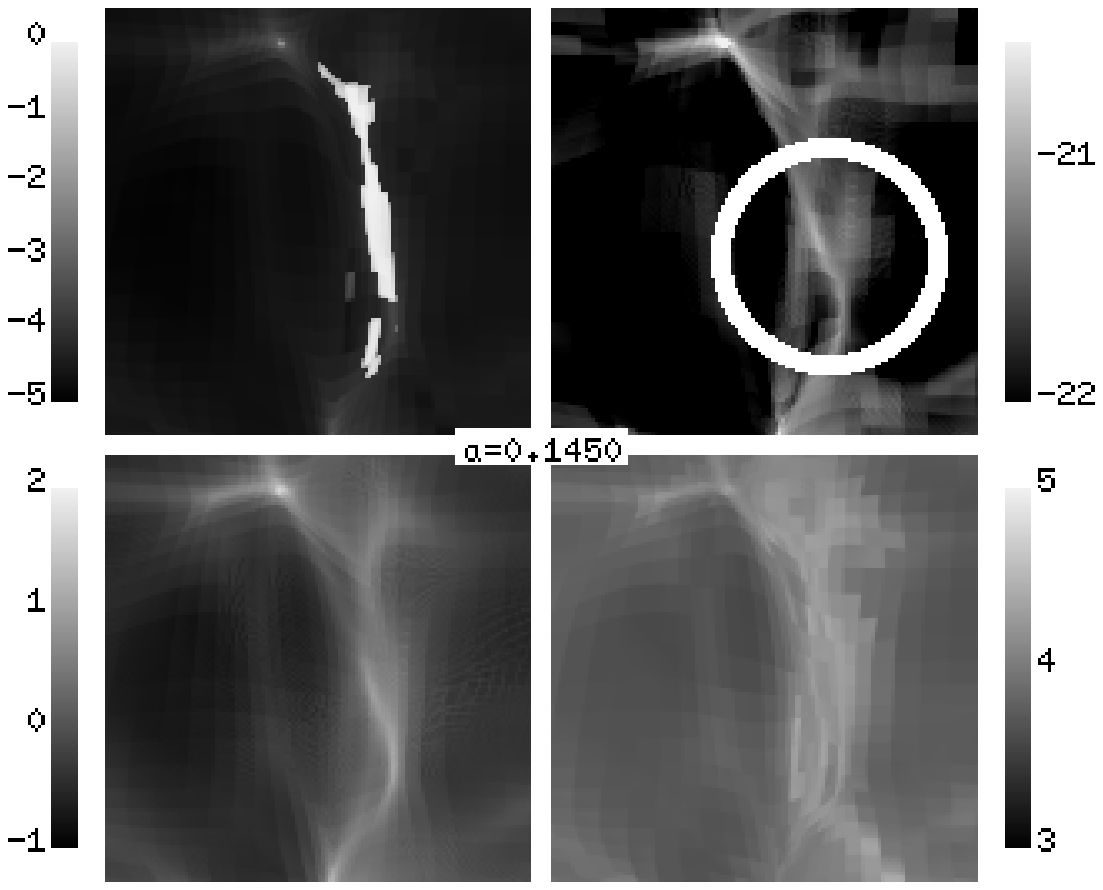}{\figdir/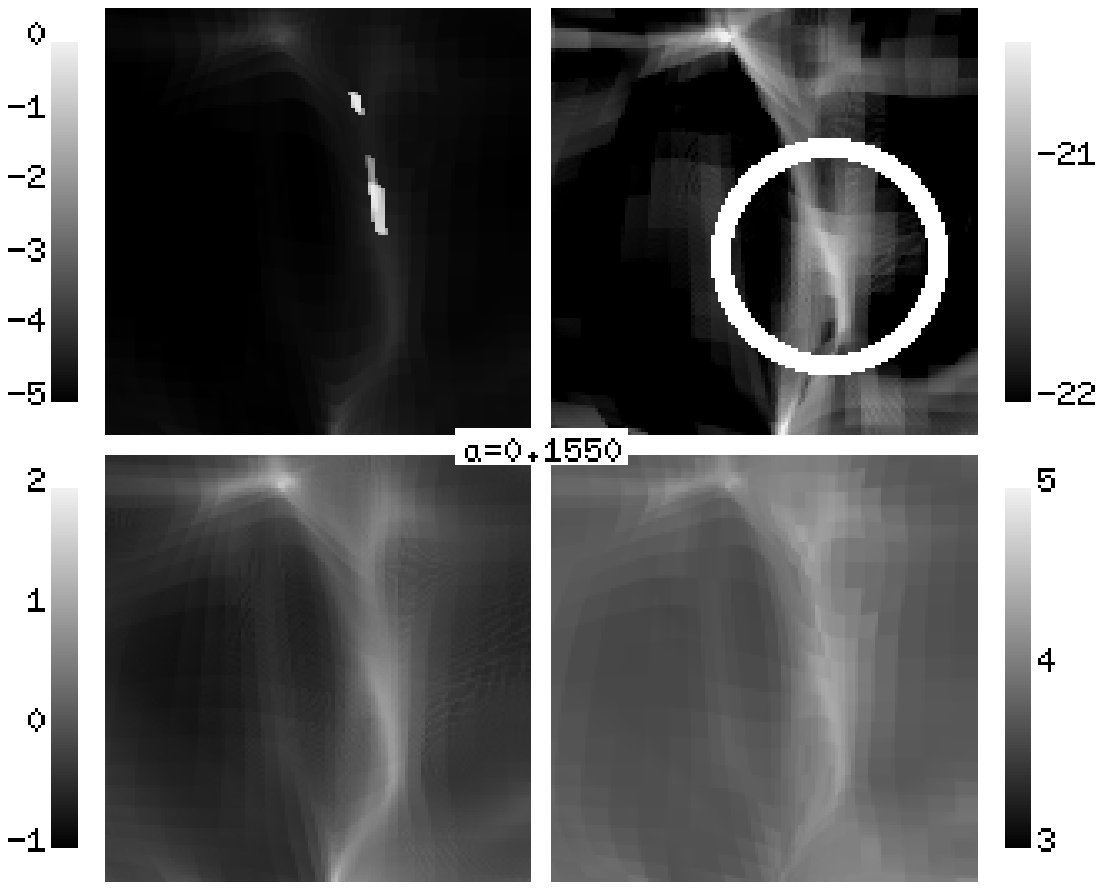}
\caption{\label{figIF}\capIF}
\end{figure}
}
The second mechanism is important when the I-front propagates across the
high density neutral filament. This is shown in Figure \ref{figSF} as a
cartoon version and further illustrated in Figure \ref{figIF} with
two snapshots from the simulation. Both, the cartoon and the snapshots,
demonstrate how this mechanism operates in the post-overlap stage, when
the low density IGM is already ionized. However, it is clear from the
cartoon that the ionization state of the surrounding low density IGM
is irrelevant, and that this mechanism works
in the pre-overlap stage as well, when the filament is embedded in the
neutral rather than ionized gas. In general, however, I-fronts propagate
much more slowly across the high density filaments than in the low density
IGM, so it is likely that the IGM around the filament is already ionized
when the I-front propagates across it.

It is important to keep in mind that the filaments are neutral not
because they are self-shielded, but because they are dense enough for
the recombination time to be shorter than the photoionization time,
i.e.\ even if filaments are neutral, they are nevertheless photoheated.
Since in the IGM with moderate overdensities ($\delta\la10$) there exists
a tight relation between the gas density and the gas temperature
(the so called ``effective equation of state'', Hui \& Gnedin 1998),
even a filament with quite moderate overdensity has a temperature
gradient pointed outward, the most efficient orientation for Biermann battery.

The first mechanism operates during the formation of a virialized
objects, i.e.\ in higher density environments than the second mechanism. This
fact explains why the field increases with density faster than
$\rho^{2/3}$, which would be the predicted law if the simple compression were the
dominant term in equation (\ref{beq}). Since high density regions collapse
first, the field per unit mass inside the protogalaxies is larger than
the field per unit mass in the lower density filaments and the IGM.

The existence of these two mechanisms is also the source of our disagreement
with Kulsrud \etal (1997), who found fields which are two orders of
magnitude less than the value we find. Since Kulsrud \etal (1997) did
not include radiative transfer in their simulations, they
were not able to follow the I-fronts, and therefore completely missed
these two important mechanisms for field generation. Their field was
therefore generated by chance misalignments between the electron density
and temperature gradients. These misalignments are also present in our
simulations, but they can be a dominant source of the magnetic field only
in the low density IGM, where none of the mechanisms described above operates.
Thus, in the low density IGM we should be finding values for the magnetic
field strength similar to the ones found by Kulsrud \etal  (1997), i.e\
of the order of $10^{-21}\dim{G}$, which is indeed the case, as can be seen
from Fig.\ \ref{figBD}.

\subsection{Primordial Field in the Protogalaxies}

As we have mentioned above, our simulations do not have enough resolution
to model the interstellar medium, and therefore cannot describe 
the field generation in the galactic ISM. Thus, the values of the magnetic field
strength that we find inside the protogalaxies should only be considered
as the truly primordial field, generated during the period of
reionization. The true field inside the protogalaxies
at, say, $z=4$, could be significantly
higher than the values we obtain (of the order of $10^{-18}\dim{G}$), although
it would be coherent only on small scales. Here, we have in mind
small scale fields generated directly by the battery
mechanism, as ionization fronts propagate through the clumpy molecular clouds
in which stars are born. However, these small scale fields
do not contribute as much to the magnetization of protogalaxies as large
scale fields do, as we show by the following rough argument.

Suppose that the battery acts on small clouds or filaments of characteristic
size $D_o$ and density $n_c$, and produces a field $B(D_o)$. If most of the
mass in the star-forming complex is in these small clumps, then when the clumps
eventually disperse their mean size will be $D\sim D_o(n_c/\langle n\rangle)^{
1/3}$, where $\langle n\rangle$ is the mean interstellar density. Because of
conservation of magnetic flux, the magnetic field in the dispersed clumps is
$B(D)\sim B(D_o)(\langle n\rangle/n_c)^{2/3}$. Space is then divided into
independent magnetic domains of size $D$, and the fields within the domains
are randomly oriented with respect to one another. Over a region
of size $l > D$, a random walk argument shows that $B(l)\sim B(D)(D/l)^{3/2}$
(Hogan 1983). In terms of the original battery field, $B(l)\sim B(D_o)(\langle
n\rangle/n_c)^{1/6}(D_o/l)^{3/2}$.

We now compare $B(l)$ computed as the sum of fields in small domains with the
coherent field $B_c(l)$ which is generated by propagation of a single 
I-front. From the scaling of the battery with system size and 
I-front velocity $v_f$ we
expect $B_c(l)\sim B(D_o)(D_o/l)[v_f(D_o)/v_f(l)]$. The ratio $B_c(l)/B(l)$
is then $B_c(l)/B(l)\sim (l/D_o)^{1/2}(n_c/\langle n\rangle)^{1/6}
[v_f(D_o)/v_f(l)]$. This expression shows that unless the ionization fronts
in the clumps propagate much more slowly than the fronts on large scales, the
coherent field dominates the superimposed incoherent fields. Although the
clumps can be very dense, which slows down the fronts, they are also much
closer to the source of ionization, which speeds them up. Therefore, 
the ionization front battery on large scales is more important for the
generation of a galactic field than the battery on the scale of molecular
clumps in star forming regions. However, if the locally generated battery
field is amplified sufficiently rapidly to equipartition with the turbulence,
it could play a role in subsequent generations of star formation in the 
protogalaxy.

Since we follow the intergalactic field with sufficient
precision, we can separate the whole problem of the generation 
of the magnetic field into two distinct phases: the formation of the primordial
field during reionization, which we can reproduce with our simulations,
and the dynamo action inside the protogalaxies, which we cannot resolve.
The latter process, though,  
is restricted to the protogalaxies only, and therefore
could be, at least to a   first approximation, considered independently
from the rest of the universe.

\def\capCA{
Evolution of the stellar mass ({\it upper-left\/}), 
the gas mass ({\it upper-right\/}), 
the rms magnetic field strength ({\it lower-left\/}), and
the absolute value of the mean $z$-component of the magnetic field of the
most massive virialized object within the $4h^{-1}\dim{Mpc}$
simulation box
({\it solid lines\/}),
the 10th most massive object
({\it long-dashed lines\/}),
the 50th most massive object
({\it dot-dashed lines\/}), and
the 250th most massive object
({\it dotted lines\/}).
}
\placefig{
\begin{figure}
\epsscale{0.70}
\insertfigure{\figdir/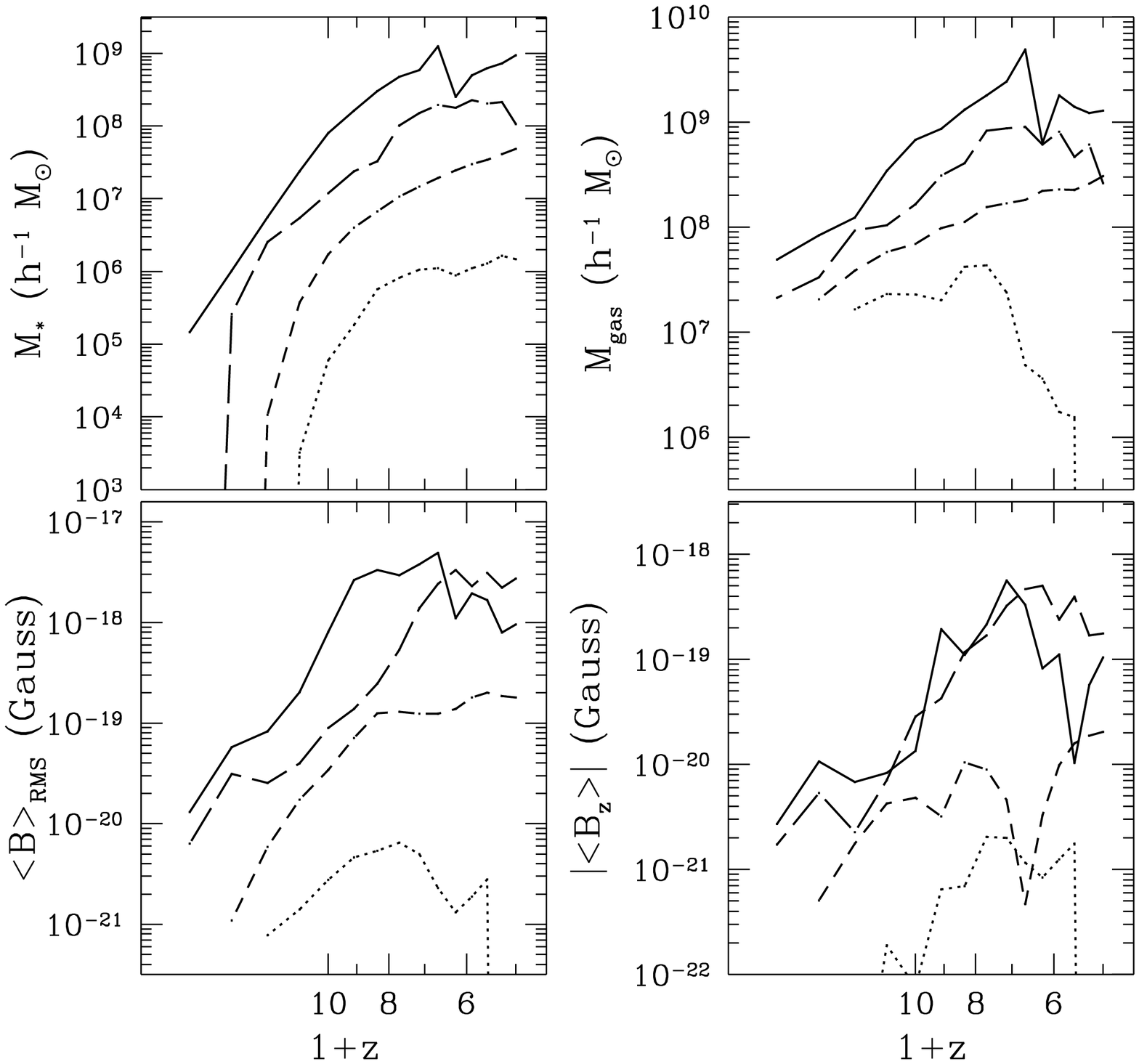}
\caption{\label{figCA}\capCA}
\end{figure}
}
Figure \ref{figCA} shows the time evolution of the stellar
mass, the gas mass, the rms magnetic field strength and 
the absolute value of the mean $z$-component of the magnetic field
($x$- and $y$- components look very similar) for four
representative objects from our simulation. We notice that the rms field
traces the evolution of the mean cosmic field from Fig.\ \ref{figEV},
and saturates at the values of about $(1-2)\times10^{-18}\dim{G}$ in the most 
massive objects. The absolute value of the $z$-component is about 10 times
smaller, implying that only 10\% of the total field is in an  ordered
state over the scale of an object\footnote{The dips in the absolute value 
of the $z$-component are due to rotation of an object, when $B_z$ passes
through zero and changes sign.}.

\def\capCC{
The final rms magnetic field strength versus baryonic mass for all virialized
objects from our simulation.
}
\placefig{
\begin{figure}
\insertfigure{\figdir/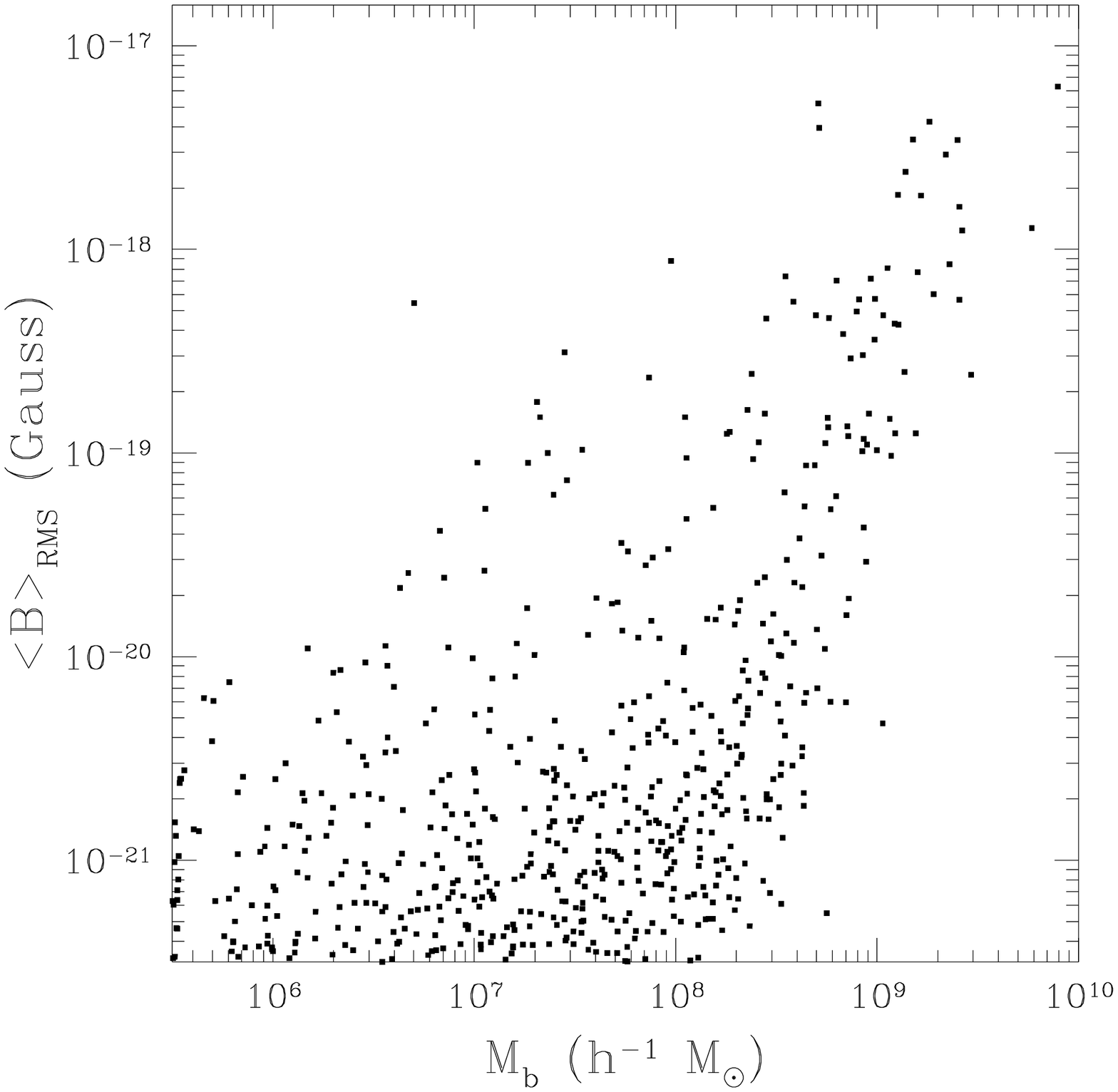}
\caption{\label{figCC}\capCC}
\end{figure}
}
Finally, in Figure \ref{figCC} we show the final (\ie saturated) value of 
the rms magnetic field strength
versus mass for all virialized objects from our simulation. The most
massive objects with baryonic mass in the range $10^9-10^{10}\dim{M}_\odot$
all have rms magnetic field strength about $10^{-18}\dim{G}$ with a factor
of a few scatter. One can also notice a correlation between the baryonic
mass and the rms magnetic field strength, albeit the scatter is quite large.
This correlation occurs for the same reason as 
the steep $B$ vs. $\rho$ dependence: the I-fronts are more efficient in
producing magnetic fields while they are breaking through the neutral
halos of the protogalaxies than when they are propagating across
filaments.

\section{Conclusions}

We have studied the generation of magnetic field during the reionization
of the universe outlining the main mechanisms,  arising in ionization fronts,
responsible for magnetogenesis. Our results allow us 
to conclude that the cosmic 
seed magnetic field  closely traces the gas density, and
that it is highly ordered on 
megaparsec scales, with a mean strength $B_0\approx 10^{-19}\dim{G}$. We find
that in protogalaxies the field is an order of magnitude or more larger than
the mean field; this is an upper limit because
unresolved small scale structure in the protogalaxies
could generate stronger fields, and amplify them rapidly.

The magnetogenesis mechanism proposed here owes its robustness to both the
relatively simple physics involved (\ie ionization fronts) and the strong
experimental evidence that reionization of the universe has indeed occurred.
Although we have assumed here that reionization is driven by stellar sources,
which appears to be consistent with available experimental data on 
the metallicity
of the IGM (Gnedin \& Ostriker 1997; Ferrara, Pettini \& Shchekinov 2000), 
quasars -- if present
prior to reionization -- are likely to produce a similar effect as they ionize the
surrounding intergalactic medium.

This field strength is about 10 orders of magnitude too small to have a
dynamical effect on galaxy formation. One way to see this is to recognize that
the Alfven speed in a medium with $n=10^{-4}\dim{cm}^{-3}$, 
$B=10^{-19}\dim{G}$ is
$2\times 10^{-6}\dim{cm}/\dim{s}$. 
This field is also too small to have much effect on
microscopic transport processes such as thermal conductivity. The relevant
parameter here is $\omega_{c}\tau_{c}$, where $\omega_c$ and $\tau_c$ are the
gyrofrequency and Coulomb collision time, respectively,
and their product must exceed unity for thermal conduction to be suppressed. 
In a plasma with $n=
10^{-4}\dim{cm}^{-3}$, 
$T=10^5\dim{K}$, $B=10^{-19}\dim{G}$, this parameter is about $3\times
10^{-3}$ for electrons. Similarly, the gyroradius of a high energy cosmic ray
propagating through this field is extremely large: a $10^{19}\dim{eV}$ 
proton in
a $10^{-19}\dim{G}$ magnetic field has a gyroradius of $10^{11}\dim{Mpc}$, 
so high
energy intergalactic cosmic rays would be essentially unmagnetized.

However, such a field, even reduced by $(1+z)^2$ from its value at formation,
might be detectable through its effects on the arrival
times of $\gamma$-rays from
extragalactic sources (Plaga 1995, Lee, Olinto, \& Sigl 1995, Waxman \& Coppi
1996). The effect of the field on the polarization of the CMB would be much
less than that modeled by Kosowsky \& Loeb (1996), who required fields
9-10 orders of magnitude larger to find a detectable magnetic signature.

This field could explain the strength currently observed in the Milky Way and
in external galaxies, provided it is further amplified by a some other process,
as for example a dynamo which has exponentiated 30 times. The protogalactic
seed fields are just barely large enough to have been amplified by a dynamo to
$\mu$G strengths by $z\sim 1$, as suggested by the observations of damped
Ly$\alpha$ systems.
 
\acknowledgements  
We thank JILA for supporting AF as a Visiting Fellow at the
time when this work was initiated. We are also happy to acknowledge support by
NASA Grant NAG 5-4063  and NSF Grant AST 98-00616 to the University of 
Colorado.
This work was also partially supported by National Computational Science
Alliance under grant AST-960015N and utilized the SGI/CRAY Origin 2000 array
at the National Center for Supercomputing Applications (NCSA).

\placefig{\end{document}}

\clearpage

\tableone

\clearpage

\newcounter{figurecap}
\setcounter{figurecap}{0}

\begin{center}
\bf Figure Captions
\end{center}

\refstepcounter{figurecap}
Fig.\ \thefigurecap---\label{figEV}\capEV

\refstepcounter{figurecap}
Fig.\ \thefigurecap---\label{figEC}\capEC

\refstepcounter{figurecap}
Fig.\ \thefigurecap---\label{figBD}\capBD

\refstepcounter{figurecap}
Fig.\ \thefigurecap---\label{figIM}\capIM

\refstepcounter{figurecap}
Fig.\ \thefigurecap---\label{figIN}\capIN

\refstepcounter{figurecap}
Fig.\ \thefigurecap---\label{figSO}\capSO

\refstepcounter{figurecap}
Fig.\ \thefigurecap---\label{figIO}\capIO

\refstepcounter{figurecap}
Fig.\ \thefigurecap---\label{figSF}\capSF

\refstepcounter{figurecap}
Fig.\ \thefigurecap---\label{figIF}\capIF

\refstepcounter{figurecap}
Fig.\ \thefigurecap---\label{figCA}\capCA

\refstepcounter{figurecap}
Fig.\ \thefigurecap---\label{figCC}\capCC

\end{document}